\documentclass[10pt]{article}
\usepackage[centertags]{amsmath}
\usepackage{amsfonts}
\usepackage[all]{xy}
\usepackage{rotating}
\newtheorem{defi}{Definition}
\newtheorem{teo}{Theorem}

\def\be{\begin{equation}}
\def\ee{\end{equation}}

\newcounter{examnum}[section]
\newcounter{remarnum}[section]
\begin{document}
\title{ Quantum Fuzzy Sets:\\ Blending Fuzzy Set Theory \\and Quantum Computation }
\author{Mirco A. Mannucci  }

\date{\today}
\maketitle
\begin{abstract}
\noindent In this article we investigate a way in which quantum
computing can be used to extend the class of fuzzy sets. The core
idea is to see states of a quantum register as characteristic
functions of quantum fuzzy subsets of a given set. As the real unit
interval is embedded in the Bloch sphere, every fuzzy set is
automatically a quantum fuzzy set. However, a generic quantum fuzzy
set can be seen as a (possibly entangled) superposition of many
fuzzy sets at once, offering new opportunities for  modeling
uncertainty. After introducing the main framework of quantum fuzzy
set theory, we analyze the standard operations of fuzzification and
defuzzification from our viewpoint. We conclude this preliminary
paper with a list of possible applications of quantum fuzzy sets to
pattern recognition, as well as future directions of pure research
in quantum fuzzy set theory.
\end{abstract}
\section{Introduction}
Fuzzy set theory (FST) has been around for four decades, producing a
remarkable bulk of theoretical and practical knowledge. Its
relevance to the real-world is beyond doubt, as many expert systems
based on fuzzy set theory and its companion, fuzzy logic (FL), have
been implemented and are currently running throughout the world. The
chief reason behind FST's remarkable success lies in its ability to
build decision systems in presence of vagueness: it provides a
suitable mathematical tool to translate properties that are
essentially ``fuzzy'' in common language usage, such as ``being quite
hot'' when talking about temperature level for a thermal device.
\\

It should be noted that, although for a given fuzzy property there
are a number of different fuzzy characteristic functions that could
express it (indeed, at least in principle, an infinite amount of
them), the choice of one or the other is left to the fuzzy expert
system designer, based on his/her understanding of the problem at
hand. It would be desirable, at least in certain cases, to be able
to represent vague predicates in more than one way {\it at once},
especially when our knowledge of the real-life system to be modeled
is insufficient to lead one way or the other.
\\

It is the contention of this paper that quantum computation can
provide the proper arena for introducing quantum fuzzy sets, i.e.
''sets'' that are in a sense superpositions of various fuzzy sets at
once. These quantum fuzzy sets can be seen as the extension of
''quantum properties'', or quantum predicates.
\\

This paper is broken down as follows: Section $1$ starts with
investigating the Logic of Quantum Computing and its relation to
Fuzzy Logic. This relation is the core ''philosophical'' motivation
behind our approach. Section $2$ shows how  standard fuzzy sets and
their operations can be represented within the quantum computation
framework. Section $3$ introduces the main topic of this paper,
quantum fuzzy sets, and sketches the way in which they add more
``room'' to FST. Section $4$ describes future directions, both
theoretical and applied, of the Quantum Fuzzy Sets program.
\\

One final word. This article is not self-contained: it assumes the
reader to be familiar with the basic ideas of Quantum Computation
(QC) and Fuzzy Set Theory. References abound: for QC the standard
one is \cite{NielsChuang}, and for fuzzy set theory (FST), two
excellent recent books covering much ground are \cite{Hajek} and
\cite{Gerla}.

\section{ From Fuzzy Logic to the Logic of Quantum Computation}
Almost since its inception, quantum mechanics has inspired new
approaches to logic. The first seminal work in this area was John
Von Neumann's proposal of a \emph{quantum logic}. Von Neumann's key
idea was that quantum mechanics may entail a radical change of view
not only of the way we understand and do physics, but of logic
itself. Quantum logic (QL) is, from an algebraic standpoint, the
orthocomplemented complete lattice of projections on a given Hilbert
space. QL has been developed for almost fifty years: a comprehensive
and well-written survey is \cite{Coecke}. A set theory based on
quantum logic has been investigated by several authors; for
instance, in the seventies and eighties Gaisi Takeuti (see
\cite{Takeuti}) proposed a quantum equivalent of boolean valued
models of set theory, thereby creating a ``universe of discourse''
where the internal logic (in the sense of categorical logic) is QL.
Later, other approaches based on more general structures, like
quantales, have been proposed (see \cite{Mulvey}).

It has been pointed out in \cite{Chiara}  that QL, with all its
undisputable merits, may  not be  the proper framework to model
quantum computation. The reason being, that QL is the {\it logic of
quantum measurements}, i.e. the logic of a physical system as it
appears to an external observer.
\\

In quantum computing measurement happens, as it were, at the very
end of the computational process. Most of the computation is the
successive application of basic unitary (and therefore reversible)
operations, known as {\it quantum gates}. To reason about this
process, it would be desirable to introduce a new type of quantum
logic, the ``internal logic'' of the processor. Indeed, something
along these lines has been proposed very recently, in \cite{Bat} and
\cite{BatZiZ}. In the two mentioned articles, a sketch for a
symmetric, para-consistent logic of internal computation is offered,
for a single-qubit and a two-qubit quantum processor. Such a logic
is quite similar in spirit to the minimalist version of linear
logic, Sambin's \emph{Basic Logic}.
\\

In the author's opinion, this internal quantum logic  (IQL for
short) is a very promising starting point, albeit incomplete. An
enhanced version that takes into account ``phase rotation''
modalities, and their geometrical content, will be presented in a
forthcoming work \cite{Mannucci}. Meanwhile, for the purpose of the
present article, we will spend a few words on the connection between
IQL and FL.
\\

Consider a one-qubit processor. Its state space is a two-dimensional
Hilbert space, and it can be mapped onto the so-called Bloch sphere.
The north pole and the south pole are the states $|0 \rangle$ and
$|1 \rangle$, respectively\footnote{We use here the standard mapping
from the qubit in normal form  $\cos(\theta) |0\rangle + e^{i \phi}
\sin( \theta) | 1\rangle $ with $ 0\leq \theta \leq \frac{\pi}{2}$
and $0\leq \phi \leq \pi$, to the point on the sphere of coordinates
$(\sin(2 \theta)\cos(\phi), \sin(\phi)\sin(2\theta), \cos(2\theta)
)$ }. The various meridians of the Bloch sphere differ by a phase.
The zero meridian is in one-to-one correspondence with the unit
interval, in a obvious way. \footnote{ The author is indebted with
Noson Yanofsky for this insight: originally, the unit interval was
realized as the vertical axis of the ``thick'' Bloch sphere, i.e. the
one containing all states, pure and impure alike. } But the unit
interval is the set of truth values of Zadeh's Fuzzy Logic. We are
thus lead to think of the Bloch sphere as {\it a generalized set of
''quantum truth values''}; once one specifies a phase, the Bloch
sphere collapses, as it were, onto a copy of the unit interval.
Indeed, the Bloch sphere is comprised of uncountable many copies of
the unit interval, just like the earth is sliced by its meridians.
\\

We wish to stress right away  that the Bloch sphere is, at least
from traditional perspectives on logic, a rather odd set of truth
values:

\begin{itemize}
  \item The Bloch sphere is not a lattice, as opposed to most
  supports for standard and not-standard logics (boolean algebras,
  Heyting algebras, distributive lattices, etc.).
  \item Truth values, i.e. points on the Bloch sphere, can differ by
  a phase, something that has no classical counterpart.
  \item The status of the pair true/false is relative. Any couple
  of  antipodal points on the Bloch sphere is a candidate for
  playing the true/false role. \footnote{An immediate consequence is that,
   upon choosing an ordered pair of antipodal points on the sphere (i.e. an observable),
    we have another  embedding of the unit interval.
    This trivial observation is nevertheless rich in consequences, as it will be shown in \cite{Mannucci}. }
\end{itemize}
As we already mentioned, an in-depth investigation on the logic of
the Bloch sphere and its higher dimensional analogues will be
tackled in another paper. For now, though, from the naive
considerations above, it should be clear that, in a sense, its logic
extends traditional FL.

Now, the step to quantum fuzzy set theory is not a big one: once one
sees the unit interval as a set of truth values, one can use it to
build ``characteristic functions''. So, why not use the
\emph{entire} Bloch sphere to do just the same?
\section{ Standard Fuzzy Sets from the Quantum Computation standpoint }

We recall that a fuzzy set is just a map:
$$f :X  \longrightarrow [0, 1]$$
from a set $X$ into the unit interval. The idea behind the
definition is simply that $f$ is a generalized characteristic
function of a subset of $X$ with unclear (fuzzy) boundaries.
\\

We are now going to describe fuzzy sets from the viewpoint of
quantum computation. In the following we are going to assume that
the set $X$ is finite, although everything we shall say would carry
through to infinite sets as well, provided that we have a quantum
machine with an infinite quantum register at our disposal.
\\

Let us assume that $X$ has cardinality $N$: $|X| = N$.  Without loss
of generality, we can think of $X$ as the set of the first $N$
natural numbers: $X=\{ 1, 2, \ldots, N \}$. We can associate to any
fuzzy characteristic function $f$ on $X$ the following quantum
state:
$$|s_{f} \rangle =  \bigotimes_{1\leq i\leq n}\left[ f(i)^{\frac{1}{2}} | 1 \rangle +
(1- f(i))^{\frac{1}{2}} |0 \rangle \right]
$$
in other words, we set each qubit in the register to the
appropriate mix given by $f$.
\begin{defi}
A state in the $N$ quantum register that can be written as
$|s_{f}\rangle$ for some $f: X\longrightarrow [0,1]$ will be
referred to as a Classical Fuzzy State (CFS).

\end{defi}
Classical Fuzzy States are exactly the ones whose components live on
the phase zero meridian of the Bloch sphere. Notice that the
standard basis of $\mathcal{C}^{2^|X|}$ becomes nothing else but the
set of characteristic functions of {\it crisp} subsets of $X$. For
instance, the state $|1100 \ldots 0\rangle$ corresponds to the
subset $\{1, 2 \}$ of $X$. This harmless observation leads to the
main insight behind this paper: states in which our quantum register
can be found encode ``characteristic functions'' (crisp or not,
classical or quantum), of a set $X$.
\\
The state $|s_{f}\rangle$ can be expanded in the standard basis as
$$|s_{f}\rangle =
f(1)^{\frac{1}{2}}f(2)^{\frac{1}{2}} \ldots f(n)^{\frac{1}{2}}
|11\ldots 1\rangle + $$ $$f(1)^{\frac{1}{2}}(1- f(2))^{\frac{1}{2}}
\ldots f(n)^{\frac{1}{2}} |10\ldots 1\rangle  +\ldots +$$ $$(1-
f(1))^{\frac{1}{2}} (1- f(2))^{\frac{1}{2}} \ldots (1-
f(n))^{\frac{1}{2}} |00\ldots 0\rangle. (\clubsuit)  $$
In other words, from the quantum computation perspective, even a
standard fuzzy set is {\it a quantum superposition of crisp sets}.
\\

Expansion of $(\clubsuit)$ above tells us a bit more, namely that an
observation of $s_{f}$ in the standard basis collapses it to (the
characteristic function of) the crisp subset  $S\subseteq X$ with
probability
$$p(S) = |\alpha_1 \alpha_2\ldots \alpha_n|^2$$
where
\[ \alpha_i=  \left\{ \begin{array}{ll}
         f(i)^{\frac{1}{2}} & \mbox{if $i \ \in   S$};\\
          (1- f(i))^{\frac{1}{2}} & \mbox{if $i  \ \not \in   S$}.\end{array} \right.   \]

One last important fact: by their very definition classical fuzzy
states are not entangled. If we want quantum magic to set in, we
must be prepared to go a bit farther.
\\

Summing up, we can say that, from our standpoint, \emph{classical
fuzzy sets are unentangled, zero-phase superpositions of crisp
sets}. Observation collapses them into one or the other of their
crisp constituents.

If we start our register in the default, ``ground state'' $|00\ldots
0 \rangle$, we can think of $f$ as the unitary map
$$U_{f} = \bigotimes U_{f(i)} $$
where
$$
U_{f(i)} =\left(%
\begin{array}{cc}
  f(i)^{\frac{1}{2}} & -(1 -  f(i))^{\frac{1}{2}}) \\
  (1 -  f(i))^{\frac{1}{2}}     & f(i)^{\frac{1}{2}} \\
\end{array}%
\right)
$$
And hence,
$$U_{f} |00 \ldots 0 \rangle = |s_{f} \rangle.$$

Each $U_{f(i)} $ does what is supposed to do, namely sending the
$i$-th qubit from the ground state $|0 \rangle$ to the proper mixing
given by $f(i)$. The geometrical interpretation is straightforward:
$U_{f(i)}$ is the rotation of the Bloch sphere around its $y$ axis
by the angle $\theta_{i}= 2 \arcsin ( f(i)^{\frac{1}{2}} )$.
\\

To show that we can indeed do fuzzy set theory from within quantum
computation, we need to be able to carry out fuzzy logical
operations on fuzzy sets. Before we embark on this step, let us
remind ourselves that FST has a large variety of alternatives
families of logical connectives (for instance, there are several
complements, several AND, etc.). We shall choose one possible set of
fuzzy operations, including logical connectives, and simply point
out that similar methods could equally well implement different
choices. Our pick is the ``probabilistic set''\footnote{ Our choice
is suggested by QC itself. As we have already seen, fuzzy
characteristic functions look like ``complex-valued probability
distributions'', from the standpoint of QC. This fact goes, at first
sight, against the general consensus that FL is {\it not}
probability logic. The contradiction is only apparent: probabilities
only occur after measurements have been taken; before, there are
only ``quantum truth values''. }:

$$\begin{tabular}{|l|r|r|}
  \hline
  complement &   $1-f(i)$\\  \hline
  intersection &     $f(i) g(i)$ \\   \hline
  \hline
\end{tabular}$$
\\
We begin with unitary connectives. The standard well-known
\textbf{NOT} gate,  $\left(
                                         \begin{array}{cc}
                                           0 & 1\\
                                           1 & 0 \\
                                         \end{array}
                                       \right)$ in the standard
                                       basis, works as desired on the
zero-phase meridian:
$$
\textbf{NOT} ( \cos (\theta) | 0 \rangle +  \sin(\theta) |1\rangle)
= \sin(\theta) |0\rangle + \cos(\theta) |1\rangle = \cos
(\frac{\pi}{2} - \theta ) | 0 \rangle +  \sin(\frac{\pi}{2} - \theta
) |1\rangle
$$
thus its action on classical fuzzy states is precisely the one given
by the complement:
$$ \textbf{NOT} (s_{f}=  \bigotimes_{i} f(i)^{\frac{1}{2}} | 1
\rangle + (1- f(i))^{\frac{1}{2}} |0 \rangle) =   \bigotimes_{i}( 1-
f(i))^{\frac{1}{2}}  | 1 \rangle + f(i)^{\frac{1}{2}} |0 \rangle.$$

As far as binary connectives, one encounters the same issues as with
standard logic gates: unitary transformations are reversible, and
thus represent only reversible gates. However, just as in the
classical logic case, there is a way around, namely adding control
qubits.
\\

As it is well known, the classical \textbf{AND} can be implemented
using the three-qubits Toffoli's Gate \textbf{T}. For the pair
$|\psi\rangle $ and $|\phi \rangle $, where $\psi, \phi \in \{ 0,1
\}$, it can be computed as follows:

$$\textbf{AND}(|\psi\rangle, |\phi \rangle ) = \textbf{T}( |\psi
\rangle \otimes |\psi \rangle \otimes |0 \rangle ) = |\psi
\rangle\otimes |\phi \rangle \otimes |\psi \phi \oplus 0\rangle$$

Here $\oplus$ denotes as usual multiplication modulo $2$. Let us now
see the action of \textbf{AND} on a pair of CFS, $|s_{f}\rangle$ and
$|s_{g}\rangle$. We shall compute it at the $i$-th component:

$$\textbf{AND}(f(i)^{\frac{1}{2}} | 1 \rangle +
(1-f(i))^{\frac{1}{2}} |0 \rangle, g(i)^{\frac{1}{2}} | 1 \rangle +
(1- g(i))^{\frac{1}{2}} |0 \rangle ) =$$
$$ f(i)^{\frac{1}{2}}g(i)^{\frac{1}{2}} T(|1\rangle, |1\rangle, | 0
\rangle) + \ f(i)^{\frac{1}{2}}(1-g(i))^{\frac{1}{2}}T(|1\rangle, 0
\rangle, |0 \rangle)  + $$ $$\
(1-f(i))^{\frac{1}{2}}g(i)^{\frac{1}{2}}T(|0\rangle, 1 \rangle, |0
\rangle)+ \ (1-f(i))^{\frac{1}{2}}(1-g(i))^{\frac{1}{2}}T(|0\rangle,
0 \rangle, |0 \rangle)  =$$
$$f(i)^{\frac{1}{2}}g(i)^{\frac{1}{2}} (|11\rangle) \ |1 \rangle + \
( f(i)^{\frac{1}{2}}(1-g(i))^{\frac{1}{2}} |10 \rangle +
 (1-f(i))^{\frac{1}{2}}g(i)^{\frac{1}{2}} |01 \rangle + $$ $$
  (1-f(i))^{\frac{1}{2}}(1-g(i))^{\frac{1}{2}} |00\rangle       )|0 \rangle $$

An observation on the third qubit of the output produces $| 1
\rangle$ with probability $f(i)g(i)$: \textbf{AND} has successfully
multiplied the amplitudes, at every $i$.
\\

Having the set-theoretical complement and intersection at our
disposal, we can as usual obtain all the fuzzy connectives by sheer
logic.
\\

We now turn our attention to {\bf fuzzification} and {\bf
defuzzification}. Let us tackle fuzzification first: what is needed
here, is an operator that takes as input a crisp value (i.e. the
register in the state where $n-1$ qubits   are set to  $|0\rangle$,
and one is set to $|1\rangle$, and fuzzifies it around the sharp
value.The exact shape of this operator depends on the desired
fuzzifier. Here, we choose a naive square-shaped fuzzifier of
diameter $k$: it creates a fuzzy value centered at the crisp one,
square-shaped, such that the base has size $2k + 1$. The value at
points surrounding the crisp value will be uniformly $\frac{1}{2}$.
Let us denote such operator as $FUZ$. If $ |\alpha_1 \dots
\alpha_n\rangle$, where
 $\alpha \ \in \ \{ 0, \ 1 \}$,  is an element of the standard basis,
here is how $FUZ$ operates on it: $FUZ (|\alpha_1 \dots
\alpha_n\rangle)=|\gamma_1,\ldots \gamma_n\rangle$
where
$$ \gamma_i = \left\{
\begin{array}{lr}
   \mbox{ $\frac{1}{\sqrt{2}}(|1\rangle + |0\rangle)$  }     &      \mbox{ if $\exists j \  |j -i| \leq k $ and $\alpha_j =1$  } \\
  0       &                                                 \mbox{ otherwise}
       \end{array} \right.
$$

$FUZ$ is linear but not unitary (fuzzification loses information).
However, we can manufacture a unitary map with the standard trick of
remembering the input vector:
$$|s_f\rangle |000\ldots 0\rangle )\longrightarrow |s_f\rangle |
U_{FUZ}( s_f)\rangle$$

Now, defuzzification. Again, as before with the AND connective, we
rely on the indirect representation of boolean maps, given by the
following recipe:

if $f: 2^{n} \rightarrow 2^{m} $ is a boolean function, we can
associate to it the unitary transformation $ U_f: 2^{n + m}
\rightarrow 2^{n + m}$ given by
$$(\spadesuit)  \ \    \forall u \in 2^n,\forall v
\in 2^m : U_f((u,v))= (u, v \oplus f(u)).$$ Before using  formula
($\spadesuit$ ) above, one must choose a classical defuzzification
operator, and discretize it (indeed, turn it into a boolean map).
Just like in the case of logical connectives, there are many choices
in FST. We shall choose as an example  the center-of-mass
defuzzification $COM: 2^n \mapsto 2^n$. If $ |\alpha_1 \dots
\alpha_n\rangle$, where $\alpha\ \in \ \{ 0, \ 1 \}$,  is a binary
string, here is how it operates: $$COM (\alpha_1 \dots
\alpha_n)=\gamma_1,\ldots \gamma_n$$ where
$$
\gamma_i = \left\{ \begin{array}{ll}
 1 &\mbox{ if $\ i = \lfloor( \frac{\sum_i g(i) i}{ \sum_i i} )\rfloor $} \\
 0 &\mbox{ otherwise}
       \end{array} \right.
$$

Now, if we want to defuzzify a state $|s_f\rangle$, we first apply
$U_{COM}$ to the state of length $2n$ $|s_f\rangle |00\ldots
0\rangle$ (i.e. $|s_f\rangle$ padded with $n$ $0$s), and then
measure the last $n$ qubits. The outcome will be a sharp
center-of-mass value, as desired (to be sure, our defuzzifier is
only a probabilistic one. It will return the center-of-mass after
repeated trials).
\\
\\
Our list is now complete. Armed with fuzzification, logical
connectives, and defuzzification, we can simulate any fuzzy
inference engine on a quantum machine, provided, of course, that a
sufficient amount of quantum memory is available.

\section{ Quantum Fuzzy Sets}
We have just seen how classical fuzzy set theory takes a rather
natural place inside Quantum Computation. It is now time to move a
step forward and introduce the main topic of the paper. Note: this
Section aims to barely sketch quantum fuzzy sets. A full treatment
of them will be tackled in \cite{Mannucci2}.

The way to quantum fuzzy sets is straightforward: we have just seen
that fuzzy characteristic functions can be represented by suitable
states of a quantum register, so why not postulate that {\it all}
states are characteristic functions of some new subsets?
\begin{defi}
A quantum fuzzy subset of a set $X$ is a point in the state space
$\mathcal{C}^{2^{|X|}}.$
\end{defi}
The reader may wonder what we have actually gained. The answer is
twofold:
\begin{itemize}
\item we can consider quantum fuzzy sets that are superpositions of
several standard fuzzy subsets, thereby creating fuzzy sets that
have different ``shapes'' at once.
\item we can create quantum fuzzy sets that are entangled
superpositions of crisp (or fuzzy) subsets. In fact, from our
perspective, \emph{every} entangled state is an entangled
superposition of crisp subsets.
\end{itemize}
If $$f_i: X \longrightarrow [0,1]$$
where $$i=1,\ldots , k$$
is a collection of fuzzy subsets of $X$, and
$$|s_{f_1}\rangle ,\ldots , |s_{f_k}\rangle$$
is the corresponding set of classical fuzzy states, one can combine
all the previous shapes  into a quantum fuzzy set via superposition:
$$|s\rangle  =c_1 |s_{f_1}\rangle  + \ldots + c_k | s_{f_k}\rangle$$
The generic quantum fuzzy state above can be entangled or not,
depending on the family of fuzzy states and the particular
amplitudes chosen.

Notice that in general classical fuzzy states are not orthogonal to
one another. Indeed, when are two CFS orthogonal to one another?
\begin{teo}
Two states $|s_{f}\rangle$, $|s_{g}\rangle$ are orthogonal to one
another, if and only if there is a $i \in X$ such that
$f(i)=0, g(i)=1$, or the other way round.
\end{teo}
The proof is immediate from expansion $\clubsuit$. An important
consequence of this fact is that the mingling of different
non-orthogonal fuzzy shapes prevent us from retrieving them sharply
at a later stage: no observable contains those shapes as its
eigenvectors.
\\

As far as entanglement goes, quantum fuzzy sets that are not
entangled can be transformed into classical fuzzy states by applying
suitable rotations of the Bloch sphere on each component (one just
rotates the meridian till one reaches zero phase). Thus, the most
interesting (and enigmatic) quantum fuzzy sets are the entangled
ones. \footnote{However, this observation does not mean that
unentangled quantum fuzzy states are trivial; far from it: their
phase may entail interesting interference phenomena when they are
combined by fuzzy connectives.}
\\

Fuzzy connectives and logical operators defined as unitary maps, as
mentioned in the previous sections, can be applied to any quantum
fuzzy set. The result is a natural  \emph{parallelization of fuzzy
operators}, out of linearity. Here is a simple example using
\textbf{NOT}:
$$\textbf{NOT} |s \rangle= \textbf{NOT}(c_1 |s_{f_1}\rangle +
\ldots +c_k |s_{f_k}\rangle) = c_1 \textbf{NOT} |s_{f_1}\rangle +
\ldots + c_k \textbf{NOT} |s_{fk}\rangle.$$

Just the same applies to all fuzzy connectives, as well as any pair
of fuzzification and defuzzification operators expressed by  unitary
transformations.
\\

We left the previous Section knowing that standard fuzzy engines can
be simulated in a quantum machine. We know more now, namely that
such inference engine accept  variables that are quantum fuzzy sets.
Such an engine can be aptly called a {\bf quantum fuzzy inference
engine} (QFE). Unlike its classical counterpart, running on a
classical machine, it can process in parallel superpositions of
classical fuzzy sets. It is thus reasonable to think of potential
applications of QFE, the topic of the last Section.

\section{Applications of Quantum Fuzzy Sets. Future Directions}
Fuzzy sets are an eminently practical tool, familiar to AI engineers
and software designers alike. Fuzzy Set Theory is generally employed
in the design of expert systems, where variable and rules allow for
vagueness. An important subclass of applications of FST is in
Pattern Recognition (PR)(see for instance \cite{fuzzyproc} and
\cite{fuzzyproce}).

The type of possible PR applications of Quantum Fuzzy Sets that we
envision exploit the features of QFS described in the last Section.
Here is a partial list:
\begin{itemize}
  \item PR applications where the training data set
  is very large (for instance, large databases of blurred images).
  The idea here is to  use the inner parallelism
  of Quantum Computing to implement a parallel fuzzy inference
  engine
  \item PR application where different features are assumed to have a
  high degree of correlation, but the exact nature of this
  correlation is unknown. Here, we would create {\it entangled}
  superposition of standard fuzzy sets, to account for new
  statistical patterns. The level of   entanglement (and therefore
  of correlation among the variables)
  could be controlled by suitable entanglement unitary
  operators.
  \item The design of entirely new PR algorithms based on Quantum
  Computing. At this initial stage, we envision algorithms based on
  {\it amplitude amplification}, like the well-known  Grover's algorithm on database search.
  An initial quantum
  fuzzy set would be initialized, representing a coarse model of the
  data. The algorithm would update the fuzzy set by amplifying some
  of its amplitudes that best fit training data. We expect that applications of
  this scheme will be relevant to fuzzy mathematical morphology
  (\cite{fuzzymorph}), as well as in the design of fuzzy filters for
  image processing (\cite{fuzzyfilter})
\end{itemize}
On the theoretical side, we see at least three independent, though
closely interrelated, lines of research:
\begin{itemize}
  \item There are several categories of Fuzzy Sets, such as Goguen's
category (\cite{GOGUEN}), and Barr's category (\cite{BARR}). In a
sequel to this work (\cite{Mannucci2}), we will introduce the
category of Quantum Fuzzy Sets,present its categorical logic,  and
investigate how the cited fuzzy sets categories are related to it.
  \item We will try to describe a general notion of quantum set, including
both quantale sets and quantum fuzzy sets into a unified framework.
  \item Lastly, there is another thread we intend to pursue,
  going back from Quantum Fuzzy Sets to Quantum Computation itself,
namely the attempt to recast quantum algorithms in the language of
QFS. From this perspective, a quantum algorithm is a series of
controlled quantum set-theoretical operations on some underlying
set. The hope is that certain known quantum algorithms will appear a
bit more intuitive when regarded from this angle, and that new
heuristic may emerge to help creating entirely ones.
\end{itemize}

 {\bf Aknowledgments} The author would like to thank Sam Lomonaco
and Ralph Wojtowicz for patiently listening to a very preliminary
version of this work and providing several good suggestions for
future directions. A special thank goes to Noson S. Yanofsky for
hints, comments, and in the editing of the first draft. All
opinions, mistakes, omissions, to be found herein, are entirely the
author's responsibility.

 \noindent Mirco A. Mannucci
\\
\\
HoloMathics, LLC\\
mirco@holomathics.com
\\
\\
Department of Computer Science, \\
George Mason University\\
mmannucc@cs.gmu.edu
\\
http://cs.gmu.edu/~mmannucc/


\begin{thebibliography}{99}

\bibitem{QuantumFuzzy} D. Aerts, T. Durt, B. Van Bogaert
{\bf  A physical example of quantum fuzzy sets and the classical
limit} Tatra Mountains Mathematical Publications 1993 , 1, pp. 5-15.
\bibitem{fuzzymorph}De Baets, B., Kerre, E., Gupta, M.{\bf The Fundamentals of
Fuzzy Mathematical Morphology: Part 1\&2}. International Journal
General Systems, vol. 23, 1995.
\bibitem{BARR} M.Barr, {\bf Fuzzy set theory and topos theory}, Canad. Math. Bull. 29 (1986), 501-508.
\bibitem{Bat}Giulia Battilotti,
{\bf Basic Logic and Quantum Computing: logical judgements by an
insider observer.} International Journal of Quantum Information $3$,
N. 1, 2005. Available at arXiv: quant-ph/0407057
 \bibitem{BatZiZ}Giulia Battilotti and P. Zizzi,
{\bf The internal logic of Bell's states.} 2004. Available at arXiv:
quant-ph/0412199
\bibitem{fuzzyproc}  H Bunke (Editor), A. Kandel (Editor), {\bf Neuro-Fuzzy Pattern Recognition}, 2001.
\bibitem{Chiara}Maria Dalla Chiara,R. Giuntini
{\bf Quantum Computational Logic: A Survey.} 2000. Available at
arXiv: quant-ph/0305029
\bibitem{Chiara1} Maria Dalla Chiara, R. Giuntini, and R Greechie
{\bf Reasoning in Quantum Theory} Kluwer, 2005.
\bibitem{Coecke} Bob Coecke, David Moore, Alexander Wilce
{\bf Operational Quantum Logic: An overview}, 2000. Available at
arXiv: quant-ph/0008019.
\bibitem{Gerla}Giangiacomo Gerla,
{\bf Fuzzy Mathematical Tools for Approximate Reasoning.} Kluwer
Academic Publishers, Dordrecht, 2001. xii+ 266 pp.
\bibitem{GOGUEN} Goguen J.A. {\bf L-fuzzy sets}. J. Math. Anal. Appl. 18 (1967), 145-157.
\bibitem{Hajek}Petr Hajek,
{\bf Metamathematics of Fuzzy Logic}  Kluwer, 1998.
\bibitem{Mannucci} Mirco A. Mannucci  {\bf Quantum Internal Logic I: A Tale of Quantum Reasoning}, in preparation
(2006).
\bibitem{Mannucci2} Mirco A. Mannucci {\bf Quantum Internal Logic II: Categories of Quantum Fuzzy Sets},
in preparation (2006).
\bibitem{Mulvey} C.J. Mulvey and J.W. Pelletier,{\bf On the Quantization of Spaces}, 2001,
 available at
 www.maths.sussex.ac.uk/Staff/CJM/
\bibitem{fuzzyfilter} M. Nachtegael, D. Van der Weken, D. Van De Ville \& E.E. Kerre (Editors)
{\bf Fuzzy Filters for Image Processing}, 2002.
\bibitem{NielsChuang}N. Nielsen, I. Chuang,
{\bf Quantum Computation and Quantum Information} Cambridge
University Press, 2000
\bibitem{fuzzyproce} Nikhil R. Pal (Editor), {\bf Pattern Recognition in Soft Computing Paradigm}, 2001.
\bibitem{rusp} E. Ruspini. {\bf Numerical Methods for Fuzzy Clustering}
Information Sciences,  1970.
\bibitem{Siler-Buckley} William Siler, James Buckley {\bf Fuzzy Expert
Systems and Fuzzy Reasoning},  Wiley, 2005.
\bibitem{Takeuti} Gaisi Takeuti,
{\bf Quantum Set Theory} Current Issues on Quantum Logic, Plenum New
York, 1983.


\end{thebibliography}
\end{document}